\documentstyle[12pt,a4wide]{article}

\renewcommand{\theequation}{\mbox{\arabic{section}.\arabic{equation}}}

\begin{document}
\author{B. Linet \thanks{E-mail: linet@celfi.phys.univ-tours.fr} \\
\small Laboratoire de Math\'ematiques et Physique Th\'eorique \\
\small CNRS/UPRES-A 6083, Universit\'e Fran\c{c}ois Rabelais \\
\small Facult\'e des Sciences et Techniques \\
\small Parc de Grandmont 37200 TOURS, France}

\title{\bf Static, massive fields and vacuum polarization potential in
Rindler space}

\date{}
\maketitle
\thispagestyle{empty}

\begin{abstract}
In Rindler space, we determine in terms of special functions 
the expression of the static, massive scalar 
or vector field generated by a point source. We find also an explicit
integral expression of the induced electrostatic potential
resulting from the vacuum polarization due to an electric charge at rest
in the Rindler coordinates. For a weak acceleration, we give then an 
approximate 
expression in the Fermi coordinates associated with the 
uniformly accelerated observer.

\end{abstract}

\section{Introduction}

In the coordinate system $(\xi^{0},\xi^{1},\xi^{2},\xi^{3})$ with $\xi^{1}>0$,
the Rindler space is described by the metric
\begin{equation}
\label{1}
ds^{2}=-(g\xi^{1} )^{2}d\xi^{0})^{2}+(d\xi^{1})^{2}+(d\xi^{2})^{2}+(d\xi^{3})^{2}
\end{equation}
where $g$ is a strictly  positive constant. Metric (\ref{1}) describes
equivalently a constant, static, homogeneous gravitational field.
There exists a horizon located at $\xi^{1}=0$.
These Rindler coordinates $(\xi^{\mu})$ 
cover only a part of the Minkowski space-time; 
they are related to the Minkowskian coordinates 
$(x^{\mu})$ for $\xi^{1}>0$ by
\begin{equation}
\label{2}
x^{0}=\xi^{1}\sinh g\xi^{0} \, ,\quad 
x^{1}=\xi^{1}\cosh g\xi^{0} \, ,\quad
x^{2}=\xi^{2} \, ,\quad x^{3}=\xi^{3} \, .
\end{equation}

In the Rindler space, the world line characterized by $\xi^{i}=\xi_{g}^{i}$ 
with
\begin{equation}
\label{2d}
\xi_{g}^{1}=\frac{1}{g}\, ,\quad \xi^{2}=0\, ,  \quad \xi^{3}=0
\end{equation}
corresponds to the world line $z^{\mu}(\tau )$ of an observer undergoing a
uniform acceleration $g$ which has the following equation in the
Minkowskian coordinates
\begin{equation}
\label{2b}
z^{0}(\tau )=\frac{1}{g}\sinh g\tau \, , \quad
z^{1}(\tau )=\frac{1}{g}\cosh g\tau \, , \quad
z^{2}(\tau )=0 \, , \quad z^{3}(\tau )=0 
\end{equation}
where $\tau$ is the proper time, the tangent vector being denoted 
$\dot{z}^{\mu}(\tau )$.
The Fermi coordinates $(y^{i})$ associated with this
accelerated observer are defined by
\begin{equation}
\label{2a}
y^{1}=\xi^{1}-\frac{1}{g}\, , \quad y^{2}=\xi^{2}\, , \quad y^{3}=\xi^{3} \, .
\end{equation}
Metric (\ref{1}) takes then the form
\begin{equation}
\label{1a}
ds^{2}=-\left( 1+2gy^{1}+g^{2}(y^{1})^{2}\right) (d\xi^{0})^{2}+(dy^{1})^{2}
+(dy^{2})^{2}+(dy^{3})^{2}
\end{equation}
allowing us to discuss the fields and the vacuum polarization
potential in the neighborhood of the line $y^{i}=0$ in an homogeneous
gravitational field.

In electromagnetism, the electromagnetic field generated by a uniformly
accelerated point charge has been a subject of considerable investigation. 
But as clearly showed by Boulware \cite{bo}, there is no problem 
in the region delimited by $\xi^{1}$
strictly positive. The field generated by a
point source having the world line (\ref{2b}) 
is calculated by using the retarded Green's function in the Minkowskian
coordinates.
By means of the transformation of coordinates (\ref{2}), this field 
can be thus expressed in
coordinates $(\xi^{0},\xi^{1},\xi^{2},\xi^{3})$ for $\xi^{1}>0$. 
This field coincides with the electrostatic potential of 
a point charge in metric (\ref{1}) which has been found by
Whittaker \cite{wh} in slightly different coordinates. It corresponds to
a charge at rest in the homogeneous gravitational field. 

The first purpose of the present paper is to give in closed form the 
expression of the
massive vector field generated by a point source having a uniform
acceleration by using the retarded Green's function
$\triangle_{R}$. So far as we know, this determination has not been done.
We will give also the result in the case of a massive scalar field
where an analogous situation exists in the region $\xi^{1}>0$ \cite{re}.

In quantum electrodynamics, when the pair creation is neglected, 
the induced current resulting from the
vacuum polarization can be determined at the first order in the fine
structure constant $\alpha$ by the Schwinger's formula \cite{sc}. 
He gave in the Minkowskian coordinates an integral expression
with the aid of the half sum of the advanced and retarded Green's functions
$\overline{\triangle}$. The second purpose of the present paper is to
show that the vacuum polarization induced by a uniformly accelerated charge
can be expressed in the Rindler space in terms of the
vector Green's function previously determined. 

The plan of the work is as follows. In Sec. 2, we write down some
preliminary formulas. We determine the massive scalar field in Sec. 3 
and the massive vector field in Sec. 4. The vacuum polarization
due to an electric charge at rest is treated in Sec. 5. We add
in Sec. 6 some concluding remarks.

\section{Preliminaries}
\setcounter{equation}{0}

In the Minkowskian coordinates, the retarded Green's function 
({\em e.g.} \cite{bog}) is
\begin{equation}
\label{31}
\triangle_{R}(x,x')=\frac{\theta (x^{0}-x^{0'})}{2\pi}\left[ 
\delta (\lambda )-\frac{m}{2\sqrt{\lambda}}\theta (\lambda )
J_{1}(m\sqrt{\lambda})\right]
\end{equation}
where $J_{1}$ is the Bessel function and the quantity $\lambda$ is given by 
\begin{equation}
\label{32}
\lambda (x,x')=(x^{0}-x^{0'})^{2}-(x^{1}-x^{1'})^{2}-(x^{2}-x^{2'})^{2}
-(x^{3}-x^{3'})^{2} \, .
\end{equation}
The half sum of the advanced and retarded Green's function is
\begin{equation}
\label{33}
\overline{\triangle}(x,x')=\frac{1}{4\pi}\left[ \delta (\lambda )
-\frac{m}{2\sqrt{\lambda}}\theta (\lambda )
J_{1}(m\sqrt{\lambda})\right] \, .
\end{equation}

We will apply these formulas when the point $x'$ coincides with the point 
$z(\tau )$ given by (\ref{2b}). By using (\ref{2}), $\lambda$ can be
expressed in the Rindler coordinates
\begin{equation}
\label{34}
\lambda (x,z(\tau ))=\frac{1}{g^{2}}\left( 2g\xi^{1}\cosh g(\xi^{0}-\tau )
-(g\xi^{1})^{2}-(g\xi^{2})^{2}-(g\xi^{3})^{2}\right) \, .
\end{equation}
The retarded time $\tau_{R}$ is defined by $\lambda (x,z(\tau_{R})=0$
and  $x^{0}-z^{0}(\tau_{R})>0$. From (\ref{34}), we find 
\begin{equation}
\label{35}
\cosh g(\xi^{0}-\tau_{R})=\frac{1+(g\xi^{1})^{2}+(g\xi^{2})^{2}
+(g\xi^{3})^{2}}{2g\xi^{1}} \quad {\rm with} \quad \xi^{0}>\tau_{R} \, .
\end{equation}
The advanced time $\tau_{A}$ is also  given also by (\ref{35}) but with
$\xi^{0}<\tau_{A}$.
In the next sections, it will be needed to know
\begin{equation}
\label{35a}
(x^{\mu}-z^{\mu}(\tau_{R})\dot{z}_{\mu}(\tau_{R})=-\xi^{1}
\sinh g(\xi^{0}-\tau_{R}) \, .
\end{equation}

We introduce the functions $x$ and $y$ of $\xi^{i}$ and $\xi_{0}^{i}$ by
\begin{eqnarray}
\label{36}
\nonumber & & x(\xi^{i},\xi_{0}^{i})=\frac{m}{2}\left[ 
\sqrt{(\xi_{0}^{1})^{2}+(\xi^{1})^{2}
+(\xi^{2}-\xi_{0}^{2})^{2}+(\xi^{3}-\xi_{0}^{3})^{2}+2\xi_{0}^{1}\xi^{1}} 
\right. \\
& & \left. -\sqrt{(\xi_{0}^{1})^{2}+(\xi^{1})^{2}
+(\xi^{2}-\xi_{0}^{2})^{2}+(\xi^{3}-\xi_{0}^{3})^{2}-2\xi_{0}^{1}
\xi^{1}} \right] \, , \\
\nonumber & & y(\xi^{i},\xi_{0}^{i})=\frac{m}{2}\left[ 
\sqrt{(\xi_{0}^{1})^{2}+(\xi^{1})^{2}
+(\xi^{2}-\xi_{0}^{2})^{2}+(\xi^{3}-\xi_{0}^{3})^{2}+2\xi_{0}^{1}\xi^{1}} 
\right. \\
\nonumber & & \left. +\sqrt{(\xi_{0}^{1})^{2}+(\xi^{1})^{2}+
(\xi^{2}-\xi_{0}^{2})^{2}+(\xi^{3}-\xi_{0}^{3})^{2}-2\xi_{0}^{1}\xi^{1}}
\right] \, .
\end{eqnarray}
We notice that $y\geq x>0$,
the equality occuring if  $\xi^{i}=\xi_{0}^{i}$.
It is easy to see that
\begin{eqnarray}
\label{38}
& & \nonumber y^{2}-x^{2}=m^{2}\sqrt{[(\xi_{0}^{1})^{2}+(\xi^{1})^{2}+(\xi^{2}-
\xi_{0}^{2})^{2}+(\xi^{3}-\xi_{0}^{3})^{2}]^{2}-4(\xi_{0}^{1}\xi^{1})^{2}} \, , \\
& & xy=m^{2}\xi^{1}\xi_{0}^{1} \, .
\end{eqnarray}

We note respectively $x_{g}$ and $y_{g}$ the functions of $\xi^{i}$ 
defined from the functions $x$ and $y$ for
the value $\xi_{0}^{i}$ given by (\ref{2d}). Hence, we can rewrite (\ref{35})
in the form
\begin{equation}
\label{40}
\ln \frac{y_{g}}{x_{g}}=\frac{x_{g}^{2}+y_{g}^{2}}{2x_{g}y_{g}}=
\cosh g(\xi^{0}-\tau_{R}) \, ,
\quad
\frac{y_{g}^{2}-x_{g}^{2}}{2xy}=\sinh g(\xi^{0}-\tau_{R}) \, .
\end{equation}
In terms of the Fermi coordinates (\ref{2a}), we have the relations
\begin{equation}
\label{41}
x_{g}y_{g}=\frac{m^{2}}{g^{2}}(1+gy^{1}) \, , \quad
y_{g}-x_{g}=m\sqrt{(y^{1})^{2}+(y^{2})^{2}+(y^{3})^{2}} \, .
\end{equation}

\section{Determination of the massive scalar field}
\setcounter{equation}{0}

The covariant equation for a massive scalar field $\psi$ in a general
metric $g_{\alpha \beta}$ is
\begin{equation}
\label{3}
\frac{1}{\sqrt{-g}}\partial_{\alpha}\left( \sqrt{-g}g^{\alpha \beta}
\partial_{\beta}\psi\right) -m^{2}\psi =-\int_{-\infty}^{\infty}
\frac{1}{\sqrt{-g}}\delta^{(4)}(x^{\lambda}-z^{\lambda}(s))ds
\end{equation}
where $z^{\lambda}(s)$ is the world line of the point source of strength unit.
In the Minkowski space-time, according to (\ref{31}) we get the 
retarded solution in the general form
\begin{equation}
\label{4}
\psi (x)=-\frac{1}{4\pi (x^{\mu}-z^{\mu}(\tau_{R}))\dot{z}_{\mu}(\tau_{R})}
-\frac{m}{4\pi}\int_{-\infty}^{\tau_{R}}
\frac{J_{1}\left( m\sqrt{\lambda (x,z(\tau ))}\right) }
{\sqrt{\lambda (x,z(\tau ))}}d\tau \, .
\end{equation}

In the case of the uniformly accelerated point source 
characterized by (\ref{2b}), formula (\ref{4}) yields
\begin{eqnarray}
\label{42a}
\nonumber & & \psi (\xi^{i})=\frac{g}{2\pi \sqrt{[1+(g\xi^{1})^{2}
+(g\xi^{2})^{2}+(g\xi^{3})^{2}]^{2}-4(g\xi^{1})^{2}}} \\
& & -\frac{mg}{4\pi}\int_{-\infty}^{\tau_{R}}\frac{J_{1}\left( m/g
\sqrt{2g\xi^{1}\cosh g(\xi^{0}-\tau )-1-(g\xi^{1})^{2}-(g\xi^{2})^{2}
-(g\xi^{3})^{2}}\right) }{\sqrt{2g\xi^{1}\cosh g(\xi^{0}-\tau )-1
-(g\xi^{1})^{2}-(g\xi^{3})^{2}}}d\tau \, .
\end{eqnarray}
The change of variable $t=g(\xi^{0}-\tau )$ in integral (\ref{42a}) gives
\begin{eqnarray}
\label{42}
\nonumber & & \psi (\xi^{i})=\frac{g}{2\pi\sqrt{[1+(g\xi^{1})^{2}
+(g\xi^{2})^{2}+(g\xi^{3})^{2}]^{2}-4(g\xi^{1})^{2}}} \\
& & +\frac{m}{4\pi}\int_{\infty}^{g(\xi^{0}-\tau_{R})}
\frac{J_{1}\left( m/g\sqrt{2g\xi^{1}\cosh t-1-(g\xi^{1})^{2}-(g\xi^{2})^{2}
-(g\xi^{3})^{2}}\right) }{\sqrt{2g\xi^{1}
\cosh t-1-(g\xi^{1})^{2}-(g\xi^{2})^{2}-(g\xi^{3})^{2}}}dt \, .
\end{eqnarray}
By considering the functions $x_{g}$ and $y_{g}$ defined in 
Sec. 2, expression (\ref{42}) takes then the form
\begin{eqnarray}
\label{45}
\nonumber & & \psi (\xi^{i})=\frac{g}{2\pi \sqrt{[1+(g\xi^{1})^{2}
+(g\xi^{2})^{2}+(g\xi^{3})^{2}]^{2}-4(g\xi^{1})^{2}}} \\
& & -\frac{m^{2}}{4\pi g}\int_{\ln y_{g}/x_{g}}^{\infty}\frac{J_{1}
\left( \sqrt{2x_{g}y_{g}
\cosh t-x_{g}^{2}-y_{g}^{2}}\right) }
{\sqrt{2x_{g}y_{g}\cosh t-x_{g}^{2}-y_{g}^{2}}}dt
\end{eqnarray}
where we have used (\ref{35}) and (\ref{40}).
The definite integral with the Bessel function appearing in (\ref{45}) 
is given by formula (\ref{a1}) of the appendix; after simplification we get
\begin{equation}
\label{46}
\psi (\xi^{i})=\frac{m^{2}}{2\pi g}\frac{x_{g}
I_{1}(x_{g})K_{0}(y_{g})+y_{g}I_{0}(x_{g})K_{1}(y_{g})}
{y_{g}^{2}-x_{g}^{2}}
\end{equation}
where $I_{\nu}$ and $K_{\nu}$ are the modified Bessel functions.

We are now in a position to determine the static scalar Green's function in
Rindler space.
The static solution to equation (\ref{3}) in metric (\ref{1}) 
for a point source located at (\ref{2d}) obeys the following equation
\begin{equation}
\label{5}
\frac{1}{\xi^{1}}\frac{\partial}{\partial \xi^{1}}\left( \xi^{1}
\frac{\partial}{\partial \xi^{1}}\psi \right)+\frac{\partial^{2}}
{\partial (\xi^{2})^{2}}\psi+\frac{\partial^{2}}
{\partial (\xi^{3})^{2}}\psi-m^{2}\psi =-\delta (\xi^{1}-\frac{1}{g})
\delta (\xi^{2})\delta (\xi^{3}) \, .
\end{equation}
The scalar Green's function $G(\xi^{i},\xi_{0}^{i})$ is the solution 
to equation (\ref{5}) with the source
\begin{equation}
\label{6}
-\frac{1}{\xi_{0}^{1}}\delta (\xi^{1}-\xi_{0}^{1})\delta (\xi^{2}-\xi_{0}^{2})
\delta (\xi^{3}-\xi_{0}^{3}) \quad {\rm with} \quad \xi_{0}^{1}>0
\end{equation}
since we must multiply by $\xi^{1}$ equation (\ref{5})  to have  a self-adjoint
operator and in consequence a Green's function 
symmetric in $\xi^{i}$
and $\xi_{0}^{i}$. So, 
we have established in closed form the expression  of the 
static scalar Green's function  in Rindler space
\begin{equation}
\label{7}
G(\xi^{i},\xi_{0}^{i})=\frac{m^{2}}{2\pi}\frac{xI_{1}(x)K_{0}(y)
+yI_{0}(x)K_{1}(y)}{y^{2}-x^{2}}
\end{equation}
where $x$ and $y$ are the functions (\ref{36}). 
When $m=0$, we obtain
\begin{equation}
\label{8}
D(\xi^{i},\xi_{0}^{i})=\frac{1}{2\pi\sqrt{[(\xi_{0}^{1})^{2}+(\xi^{1})^{2}
+(\xi^{2}-\xi_{0}^{2})^{2}+(\xi^{3}-\xi_{0}^{3})^{2}]^{2}
-4(\xi_{0}^{1}\xi^{1})^{2}}} \, .
\end{equation}
The Green's function (\ref{7}) is well defined at the horizon $\xi^{1}=0$.

\section{Determination of the massive vector field}
\setcounter{equation}{0}

The Proca  equations in covariant form for a massive vector field $A^{\mu}$ 
in a general metric $g_{\alpha \beta}$ are
\begin{eqnarray}
\label{9}
\nonumber & & \frac{1}{\sqrt{-g}}\partial_{\alpha}\left[ 
\sqrt{-g}g^{\alpha \beta}
g^{\gamma \delta}(\partial_{\beta}A_{\gamma}-\partial_{\gamma}A_{\beta})\right]
-m^{2}A^{\delta}=\int_{-\infty}^{\infty}\dot{z}^{\delta}(s)
\frac{1}{\sqrt{-g}}\delta^{(4)}(x^{\lambda}-z^{\lambda}(s))ds \\
& & {\rm and} \quad \partial_{\mu}(\sqrt{-g}A^{\mu})=0
\end{eqnarray}
where $z^{\lambda}(s)$ is the world line of the point source of strength unit.

According to (\ref{31}), we obtain the expressions of the Minkowskian
components $A^{\mu}$ in the general form
\begin{equation}
\label{51}
A^{\mu}(x)=\frac{\dot{z}^{\mu}(\tau_{R})}{4\pi (x^{\mu}-z^{\mu}(\tau_{R}))
\dot{z}_{\mu}(\tau_{R})}+\frac{m}{4\pi}\int_{-\infty}^{\tau_{R}}
\dot{z}^{\mu}(\tau )\frac{J_{1}\left( m\sqrt{\lambda (x,z(\tau ))}\right) }
{\sqrt{\lambda (x,z(\tau ))}}d\tau
 \, .
\end{equation}
We can now express  Minkowskian components (\ref{51}) in the 
Rindler coordinates
\begin{eqnarray}
\label {52}
\nonumber & & A^{0}(\xi^{0},\xi^{i})=-\frac{g\cosh g\tau_{R}}{2\pi
\sqrt{[1+(g\xi^{1})^{2}
+(g\xi^{2})^{2}+(g\xi^{3})^{2}]^{2}-4(g\xi^{1})^{2}}} \\
& & +\frac{mg}{4\pi}\int_{-\infty}^{\tau_{R}}\cosh g\tau \\ 
\nonumber & & \times \frac{J_{1}\left( m/g\sqrt{2g\xi^{1}
\cosh g(\xi^{0}-\tau )
-1-(g\xi^{1})^{2}
-(g\xi^{2})^{2}-(g\xi^{3})^{2}}\right) }
{\sqrt{2g\xi^{1}\cosh g(\xi^{0}-\tau )
-1-(g\xi^{1})^{2}-(g\xi^{2})^{2}-(g\xi^{3})^{2}}}d\tau \, , 
\end{eqnarray}
\begin{eqnarray}
\label{52a}
\nonumber & & A^{1}(\xi^{0},\xi^{i})=-\frac{g\sinh g\tau_{R}}{2\pi
\sqrt{[1+(g\xi^{1})^{2}
+(g\xi^{2})^{2}+(g\xi^{3})^{2}]^{2}-4(g\xi^{1})^{2}}} \\
& & +\frac{mg}{4\pi}\int_{-\infty}^{\tau_{R}}\sinh g\tau \\ 
\nonumber & & \times \frac{J_{1}\left( m/g 
\sqrt{2g\xi^{1}\cosh g(\xi^{0}-\tau )
-1-(g\xi^{1})^{2}
-(g\xi^{2})^{2}-(g\xi^{3})^{2}}\right)}
{\sqrt{2g\xi^{1}\cosh g(\xi^{0}-\tau )
-1-(g\xi^{1})^{2}-(g\xi^{2})^{2}-(g\xi^{3})^{2}}}d\tau
\end{eqnarray}
the other components vanishing. By means of the transformation of
coordinates (\ref{2}), we obtain the components of the vector field
in the Rindler coordinates
\begin{equation}
\label{53}
A_{\xi^{0}}=-g\xi^{1}\cosh g\xi^{0}A^{0}+g\xi^{1}\sinh g\xi^{0}A^{1}\, , \quad
A_{\xi^{1}}=-\sinh g\xi^{0}A^{0}+\cosh g\xi^{0}A^{1} \, .
\end{equation}
From (\ref{53}) the component $A_{\xi^{0}}$ can be written
\begin{eqnarray}
\label{54}
\nonumber & & A_{\xi^{0}}(\xi^{i})=\frac{g^{2}\xi^{1}\cosh g(\xi^{0}-\tau_{R})}
{2\pi\sqrt{[1+(g\xi^{1})^{2}+(g\xi^{2})^{2}+(g\xi^{3})^{2}]^{2}-
4(g\xi^{1})^{2}}} \\
& &-\frac{mg^{2}\xi^{1}}{4\pi}\int_{-\infty}^{\tau_{R}}
\cosh g(\xi^{0}-\tau ) \\
& & \nonumber \times \frac{J_{1}\left( m/g
\sqrt{2g\xi^{1}\cosh g(\xi^{0}-\tau )
-1-(g\xi^{1})^{2}-(g\xi^{2})^{2}-(g\xi^{3})^{2}}\right) }
{\sqrt{2g\xi^{1}\cosh
g(\xi^{0}-\tau )-1-(g\xi^{1})^{2}-(g\xi^{2})^{2}-(g\xi^{3})^{2}}}d\tau
\end{eqnarray}
and the component $A_{\xi^{1}}$
\begin{eqnarray}
\label{55}
\nonumber & & A_{\xi^{1}}(\xi^{i})=\frac{g\sinh g(\xi^{0}-\tau_{R})}
{2\pi\sqrt{[1+(g\xi^{1})^{2}+(g\xi^{2})^{2}+(g\xi^{3})^{2}]^{2}
-4(g\xi^{1})^{2}}} \\
& & -\frac{mg}{4\pi}\int_{-\infty}^{\tau_{R}}
\sinh g(\xi^{0}-\tau ) \\
& & \nonumber \times \frac{J_{1}\left( m/g\sqrt{2g\xi^{1}
\cosh g(\xi^{0}-\tau )
-1-(g\xi^{1})^{2}-(g\xi^{2})^{2}-(g\xi^{3})^{2}}\right) }
{\sqrt{2g\xi^{1}\cosh
g(\xi^{0}-\tau )-1-(g\xi^{1})^{2}-(g\xi^{2})^{2}-(g\xi^{3})^{2}}}d\tau \, .
\end{eqnarray}
By introducing the functions $x_{g}$ and $y_{g}$ defined in Sec. 2, 
we rewrite (\ref{54}) in the form
\begin{eqnarray}
\label{56}
\nonumber & & A_{\xi^{0}}(\xi^{i})=\frac{g[1+(g\xi^{1})^{2}+(g\xi^{2})^{2}+
(g\xi^{3})^{2}]}{4\pi\sqrt{[1+(g\xi^{1})^{2}+(g\xi^{})^{2}+(g\xi^{3})^{2}]
-4(g\xi^{1})^{2}}} \\
& & -\frac{m^{2}\xi^{1}}{4\pi}\int_{\ln y_{g}/x_{g}}^{\infty}
\frac{J_{1}\left( \sqrt{2x_{g}y_{g}\cosh t-x_{g}^{2}-y_{g}^{2}}\right) }
{\sqrt{2x_{g}y_{g}\cosh t-x_{g}^{2}-y_{g}^{2}}}\cosh tdt \, .
\end{eqnarray}
From formula (\ref{a2}) of the appendix, component (\ref{56})
has then the explicit expression 
\begin{equation}
\label{57}
A_{\xi^{0}}(\xi^{i})=\frac{m^{2}\xi^{1}}{2\pi}
\frac{y_{g}I_{1}(x_{g})K_{0}(y_{g})
+x_{g}I_{0}(x_{g})K_{1}(y_{g})}{y_{g}^{2}-x_{g}^{2}} \, .
\end{equation}
On the other hand with the help of formula (\ref{a3}) of the appendix, it is
easy to see that $A_{\xi^{1}}=0$.

We are now in a position to determine the static vector Green's function in
Rindler space. 
In the static case for a point source located at (\ref{2d}),
equations (\ref{9}) reduce to one equation for the only non-zero
component $A_{\xi^{0}}$ 
\begin{equation}
\label{10}
\xi^{1}\frac{\partial}{\partial \xi^{1}}\left( \frac{1}{\xi^{1}}
\frac{\partial}{\partial \xi^{1}}A_{\xi^{0}}\right) +\frac{\partial^{2}}
{\partial (\xi^{2})^{2}}A_{\xi^{0}}+\frac{\partial^{2}}
{\partial (\xi^{3})^{2}}A_{\xi^{0}}-m^{2}A_{\xi^{0}}=-\delta (\xi^{1}-
\frac{1}{g})\delta (\xi^{2})\delta (\xi^{3}) 
\end{equation}
whose the solution is given by (\ref{57}).
The source of equation (\ref{10}) for the static vector Green's function
$G_{\xi^{0}}(\xi^{i},\xi_{0}^{i})$ is
\begin{equation}
\label{11}
-\xi_{0}^{1}\delta (\xi^{1}-\xi_{0}^{1}\delta (\xi^{2}-\xi_{0}^{2})
\delta (\xi^{3}-\xi_{0}^{3}) \quad {\rm with} \quad \xi_{0}^{1}>0
\end{equation}
because we must divide by $\xi^{1}$ equation (\ref{10}) to have  a self-adjoint
operator and to obtain a Green's function symmetric in $\xi^{i}$
and $\xi_{0}^{i}$. Therefore we have
\begin{equation}
\label{12}
G_{\xi^{0}}(\xi^{i},\xi_{0}^{i})=\frac{m^{2}\xi^{1}\xi_{0}^{1}}{2\pi}
\frac{yI_{1}(x)K_{0}(y)+xI_{0}(x)K_{1}(y)}{y^{2}-x^{2}}
\end{equation}
where $x$ and $y$ are functions (\ref{36}). When $m=0$, {\em i.e.}
in electrostatics, the massless Green's function is
\begin{equation}
\label{12a}
D_{\xi^{0}}(\xi^{i},\xi_{0}^{i})=\frac{(\xi_{0}^{1})^{2}+(\xi^{1})^{2}
+(\xi^{2}-\xi_{0}^{2})^{2}+(\xi^{3}-\xi_{0}^{3})^{2}}{4\pi
\sqrt{[(\xi_{0}^{1})^{2}+(\xi^{1})^{2}+(\xi^{2}-\xi_{0}^{2})^{2}
+(\xi^{3}-\xi_{0}^{3})^{2}]^{2}-4(\xi_{0}^{1}\xi^{1})^{2}}} \, .
\end{equation}

When $\xi^{1}\rightarrow 0$, expression  (\ref{12}) has the asymptotic form
\begin{eqnarray}
\label{13}
\nonumber & & G_{\xi^{0}}(\xi^{i},\xi_{0}^{i})
\sim \frac{(\xi_{0}^{1}\xi^{1})^{2}}
{2\pi [(\xi_{0}^{1})^{2}
+(\xi^{2}-\xi_{0}^{2})^{2}+(\xi^{3}-\xi_{0}^{3})^{2}]}  \\
& & \left[ +\frac{1}{2}K_{0}\left(
m\sqrt{(\xi_{0}^{1})^{2}+(\xi^{2}-\xi_{0}^{2})^{2}+
(\xi^{3}-\xi_{0}^{3})^{2}}\right) \right. \\ 
\nonumber & & \left. +\frac{m}
{\sqrt{(\xi_{0}^{1})^{2}+(\xi^{2}-\xi_{0}^{2})^{2}+
(\xi^{3}-\xi_{0}^{3})^{2}}}K_{1}\left(
m\sqrt{(\xi_{0}^{1})^{2}+(\xi^{2}-\xi_{0}^{2})^{2}+
(\xi^{3}-\xi_{0}^{3})^{2}}\right) \right] 
\end{eqnarray}  
since $I_{0}(0)=0$ and $I_{1}(x)\sim x/2$ as $x\rightarrow 0$.

So, we have established in closed form  
expression (\ref{12}) of the static vector Green's function in 
Rindler space. By virtue of (\ref{13}), the scalar invariant 
of the Proca field $A_{\mu}A^{\mu}$
in metric (\ref{1}) has a regular behavior at the
horizon $\xi^{1}=0$,
likewise the associated field 
$F_{\mu \nu}=\partial_{\mu}A_{\nu}-\partial_{\nu}A_{\mu}$.

\section{Vacuum polarization potential of a charge}
\setcounter{equation}{0}

We consider now a uniformly accelerated electric charge $e$ 
in the Minkowski space-time.
We denote $a^{\mu}$ the vector potential generated by this charge. The 
Maxwell equations are equivalent to
\begin{equation}
\label{21}
\Box a^{\mu}=j^{\mu} \quad {\rm and} \quad \partial_{\mu}a^{\mu}=0
\end{equation}
where $j^{\mu}$ is the conserved electric current associated with this charge. 
The Proca equations (\ref{9}) with $m=0$ reduce to (\ref{21}).
The induced vector potential $<a^{\mu}>$ resulting from the vacuum
polarization can be derived from the Schwinger formula \cite{sc}
\begin{eqnarray}
\label{22}
\nonumber & & <a^{\mu}(x)>=-\frac{4\alpha e}{\pi}\int \int_{0}^{1}
\overline{\triangle}\left[ \frac{2}{\sqrt{1-v^{2}}}(x-x')\right] \\ 
& & \times \frac{1-v^{2}/3}{(1-v^{2})^{2}}v^{2}j^{\mu}(x')d^{4}x' dv
\end{eqnarray}
where $\overline{\triangle}$ is given by (\ref{33}), $m$ being the mass
of the electric charge.

We can apply formula (\ref{22}) for a uniformly accelerated electric charge.
Taking into account the specific property of the advanced time $\tau_{A}$
in the present case, the field determined with $\overline{\triangle}$
coincides with the one calculated with $\triangle_{R}$. 
We proceed as in Sec. 4 and from (\ref{57}) we obtain 
\begin{eqnarray}
\label{23}
\nonumber & & <a_{\xi^{0}}(\xi^{i})>=\frac{4\alpha e}{\pi}
\frac{m^{2}\xi^{1}}{2\pi}\int_{0}^{1} \\
& & \left[ y_{g}I_{1}\left( \frac{2x_{g}}{\sqrt{1-v^{2}}}\right) 
K_{0}\left( \frac{2y_{g}}{\sqrt{1-v^{2}}}\right)
+x_{g}I_{0}\left( \frac{2x_{g}}{\sqrt{1-v^{2}}}\right)
K_{1}\left( \frac{2y_{g}}{\sqrt{1-v^{2}}}\right) \right] \\
\nonumber & & \times \frac{1}{y_{g}^{2}-x_{g}^{2}}
\frac{\sqrt{1-v^{2}}}{2}\frac{1-v^{2}/3}{(1-v^{2})^{2}}v^{2}dv
\end{eqnarray}
and also $<a_{\xi^{1}}>=0$.

Now, we can express formula (\ref{23}) with the aid of the static vector
Green's function (\ref{12}) under the form
\begin{equation}
\label{25}
<a_{\xi^{0}}(\xi^{i})>=g\frac{\alpha e}{\pi}\int_{0}^{1}
G_{\xi^{0}}\left[ \frac{2}{\sqrt{1-v^{2}}}\xi^{i},\frac{2}
{\sqrt{1-v^{2}}}\xi_{0}^{i} \right] \frac{1-v^{2}/3}{1-v^{2}}v^{2}dv \, .
\end{equation}
We have already found 
result (\ref{25}) without rigour in our previous work 
\cite{li} by putting forward 
 a development in power series in $1/m^{2}$ of the Schwinger
formula (\ref{22}) in the Minkowskian coordinates. 
However in the present work, we have now 
an explicit expression for $G_{\xi^{0}}$.

Integral expression (\ref{23}) can be approximatively 
evaluated if we suppose that
the points $\xi^{i}$ and $\xi_{0}^{i}$ are such  that $x_{g}\gg 1$ and
$y_{g}\gg 1$ with $y>x$. This means that
\begin{equation}
\label{26}
g\ll m \quad {\rm and} \quad \mid \xi^{i}-\xi_{g}^{i}\mid \ll \frac{1}{g} \, .
\end{equation} 
We treat only weak accelerations and we remain 
in the neighborhood of the world line of
the charge. Taking into account the asymptotic behavior
of the modified Bessel functions
\[
I_{\nu}(z)= \frac{1}{\sqrt{2\pi z}}{\rm e}^{z}\left( 1+O(\frac{1}{z})\right)\, , 
\quad
K_{\nu}(z)= \sqrt{\frac{\pi}{2z}} {\rm e}^{-z}
\left( 1+O(\frac{1}{z})\right) \, ,
\]
integral (\ref{23}) has the approximate expression
\begin{equation}
\label{28}
<a_{\xi^{0}}(\xi^{i})>\approx \frac{4\alpha e}{\pi} 
\frac{m^{2}\xi^{1}}{16\pi
\sqrt{x_{g}y_{g}}(y_{g}-x_{g})} \int_{0}^{1}
\exp \left( \frac{2}{\sqrt{1-v^{2}}} (x_{g}-y_{g})\right)
\frac{1-v^{2}/3}{1-v^{2}} v^{2}dv \, .
\end{equation}
We recognize in (\ref{28}) the Uehling potential $U(mr)/4\pi r$ \cite{ue}
which is the vacuum polarization potential of  a fixed charge 
in the Minkowski space-time 
\begin{equation}
\label{29}
U(mr)=\frac{\alpha}{\pi}\int_{0}^{1}\exp \left( -\frac{2mr}{\sqrt{1-v^{2}}}
\right)\frac{1-v^{2}/3}{1-v^{2}}v^{2}dv \, ;
\end{equation}
it can be expressed in terms of special functions. Because (\ref{26})
the Fermi coordinates (\ref{2a}) are well adapted to this case with 
$\mid y^{i}\mid \ll 1/g$.
By using (\ref{41}), we can express (\ref{28}) in the form
\begin{equation}
\label{30}
<a_{\xi^{0}}(y^{i})>\approx \frac{e\sqrt{1+gy^{1}}}
{4\pi\sqrt{(y^{1})^{2}+(y^{2})^{2}+
(y^{3})^{2}}}U\left( m\sqrt{(y^{1})^{2}+(y^{2})^{2}+(y^{3})^{2}}\right) \, .
\end{equation}
By neglecting terms in $\mid gy^{i}\mid^{2}$ in (\ref{30}), we finally get
\begin{equation}
\label{30a}
<a_{\xi^{0}}(y^{i})>\approx \frac{e}
{4\pi\sqrt{(y^{1})^{2}+(y^{2})^{2}+(y^{3})^{2}}}
\left( 1+\frac{1}{2}gy^{1} \right)
U\left( m\sqrt{(y^{1})^{2}+(y^{2})^{2}+(y^{3})^{2}})\right)
\end{equation}
where $U$ is given by (\ref{29}).
 
We now return to the explicit expression (\ref{23}) and  we can discuss the 
regularity at the horizon $\xi^{1}=0$. From asymptotic form (\ref{13}), we see
immediately that the induced electrostatic potential is
proportional to $(\xi^{1})^{2}$ as $\xi^{1}\rightarrow \infty$.
The induced charge density is derived via the Maxwell equations
which reduce in the static case to
\begin{equation}
\label{21a}
<j_{\xi^{0}}>=\left[ \xi^{1}\frac{\partial}{\partial \xi^{1}}
\left( \frac{1}{\xi^{1}}\frac{\partial}{\partial \xi^{1}}\right)
+\frac{\partial^{2}}{\partial (\xi^{2})^{2}}
+\frac{\partial^{2}}{\partial (\xi^{3})^{2}}\right] <a_{\xi^{0}}> \, .
\end{equation}
So, the induced charge density (\ref{21a}) is proportional to $\xi^{1}$ as
$\xi^{1}\rightarrow 0$.

\section{Conclusion}
\setcounter{equation}{0}

Probably due to the lack of physical motivation, the expression in terms of
special functions of the static, massive scalar or vector field 
generated by a point source at rest in the Rindler metric (\ref{1}) 
had not been determined.
We have filled up this gap by giving formulas (\ref{46}) and (\ref{57}) 
and also the Green's functions (\ref{7}) and (\ref{12}).

Furthermore, our method for determining these fields allows us to treat the
vacuum polarization for a uniformly accelerated electric charge by using
the Schwinger formula. In the Rindler coordinates,
we have found the induced electrostatic potential as an explicit
integral expression (\ref{23}). In the Fermi coordinates $(y^{i})$ 
and with the 
assumption of a weak acceleration $g$, we have derived approximate
expression (\ref{30a}) which is just the Uehling potential multiplied by
$1+gy^{1}/2$. This potential corresponds equivalently to the vacuum 
polarization potential of an
electric charge at rest in the homogeneous gravitational field
described by metric (\ref{1a}).

\section*{Appendix}
\renewcommand{\thesection}{\mbox{A}}
\renewcommand{\theequation}{\mbox{A.\arabic{equation}}}
\setcounter{equation}{0}

The basic formula on the Bessel function $J_{0}$ which is necessary in our 
present work is the following definite integral \cite{le}
\begin{equation}
\label{a0}
\int_{\ln y/x}^{\infty}J_{0}\left( 
\sqrt{2xy\cosh t-x^{2}-y^{2}}\right) dt=2I_{0}(x)K_{0}(y)
\end{equation}
for $y>x>0$, where $I_{\nu}$ and $K_{\nu}$ are the modified Bessel functions.
Now, we derive (\ref{a0}) with respect to $x$ and $y$ and we combine 
these results in order to obtain the two following relations 
\begin{eqnarray}
\label{a1}
\nonumber & & \int_{\ln y/x}^{\infty}\frac{J_{1}\left( \sqrt{2xy\cosh t-x^{2}
-y^{2}}\right) }{\sqrt{2xy\cosh t-x^{2}-y^{2}}}dt= \\
& & \frac{2}{y^{2}-x^{2}}-\frac{2(xI_{1}(x)K_{0}(y)+yI_{0}(x)K_{1}(y))}
{y^{2}-x^{2}} \, ,
\end{eqnarray}
\begin{eqnarray}
\label{a2}
\nonumber & & \int_{\ln y/x}^{\infty}\frac{J_{1}\left( \sqrt{2xy\cosh t-x^{2}
-y^{2}}\right) }{\sqrt{2xy\cosh t-x^{2}-y^{2}}}\cosh t dt= \\
& & \frac{x^{2}+y^{2}}{xy(y^{2}-x^{2})}-\frac{2(yI_{1}(x)K_{0}(y)
+xI_{0}(x)K_{1}(y))}{y^{2}-x^{2}}
\end{eqnarray}
where we have used $J_{0}(0)=1$, $J_{0}'=-J_{1}$, $I_{0}'=I_{1}$ and
$K_{0}'=-K_{1}$.

By noticing that
\[
\int_{0}^{\infty}J_{1}(au)du=\frac{1}{a} \, ,
\]
it is easy to prove 
\begin{equation}
\label{a3}
\int_{\ln y/x}^{\infty}\frac{J_{1}\left( \sqrt{2xy\cosh t-x^{2}-y^{2}}\right) }
{\sqrt{2xy\cosh t-x^{2}-y^{2}}}\sinh t dt=\frac{1}{xy} \, .
\end{equation}

\end{document}